\documentclass[11pt]{cernrep}
\usepackage{graphicx}

\begin{document}

\def\ihat{\hat\imath}\def\Tr{\rm Tr\,}
\def\tr{\rm tr \,}
\def\Eq#1{Eq.~(\ref{#1})}
\def\eff{\rm eff}
\def\MEV{\rm MeV}
\def\GEV{\rm GeV}
\def\<{\langle}
\def\>{\rangle}
\def\MF{mean field}

\title{BULK THERMODYNAMICS OF $SU(N)$ LATTICE GAUGE THEORIES AT LARGE-$N$}

\author{Barak Bringoltz}

\institute{Rudolf Peierls Centre for Theoretical Physics,\\
        University  of  Oxford,\\
        1 Keble Road, Oxford, OX1 3NP, UK}

\maketitle

\begin{abstract}

We present a study of bulk thermodynamical quantities in the
deconfined phase of pure lattice $SU(N)$ gauge theories. We find
that the deficit in pressure and entropy with respect to their
free-gas values, for $N=4,8$, is remarkably close to that of $SU(3)$. This
suggests that understanding the strongly interacting nature of the
deconfined phase, which is crucial for RHIC physics, can be done at
large $N$. There, different analytical approaches simplify or become
soluble, and one can check their predictions and point to
their important ingredients.

\end{abstract}

\section{Introduction}
\label{sec:pressure}

Recent discoveries at RHIC point to the inadequacy of using a weakly
coupled plasma
description for the deconfined phase of QCD (see for example
\cite{Heinz} and references therein). Instead, properties of
collective flow, can be described very well with ideal
hydrodynamics, which suggests that the phase is in fact strongly
interacting. Related to these are theoretical studies in the
framework of lattice QCD, which found signs of these strong interactions.
These include a large deficit in thermodynamical bulk
properties, such as pressure, entropy, and energy density, with
respect to their free gas values, up to temperatures of a few times
the transition temperature $T_c$ \cite{Boyd_et_al} \footnote{An additional evidence
for a `strongly coupled plasma' is
  the survival of hadronic states above $T_c$ \cite{Petreczky:2004xs}.}. These deficits were
found in the pure $SU(3)$ gauge theory, as well as with different
types of fermions. The deficit in all these cases is quite similar,
and is hard to reproduce with standard perturbation theory. This is in
contrast to the fermionic addition to the pressure at a small,
nonzero, chemical potential, which is explained very well by
perturbation theory (for example \cite{Philippe} and references thererin).

The analytical approaches used to study this phenomenon include
quasi-particle models, Polyakov loop models, variants
of perturbation theory, models of loosely bound
states, gravity dual models, and also perturbative small
volume calculations of the large-$N$ gauge theory\footnote{An list of
  some references for these analytical efforts appears
  in \cite{pressure_paper}.}.
 Some of these approaches simplify or even become soluble at the `t Hooft limit of
 large-$N$, which is a free theory in the confined phase (at least in
 the mesonic sector), and consists of only planar diagrams in a
 perturbative expansion of the deconfined phase. An additional
 important point here is that, provided the chiral limit is taken after the
 large-$N$ limit, the planar pure gauge theory is effectively
 quenched (for example see \cite{NN}).
 This makes any calculation in the pure gauge case relevant for the physical case, that includes
 fermions.

The simplifications of the planar limit, together with the possibility
that the pressure deficit is `hidden' in the gluonic sector, and the
need to check and constraint the analytical explanations, lead us to
study bulk thermodynamical properties of $SU(4)$, and $SU(8)$ pure
lattice gauge theories and compare to the study by Boyd et al. of $SU(3)$ \cite{Boyd_et_al}. Recent calculations
of various  properties of $SU(N)$ gauge theories
\cite{Lucini:2002ku}--\cite{Lucini:2004my}
have demonstrated that $N=8$ is very close to $N=\infty$
for most purposes. These also provide information on the transition coupling for various
lattice sizes and $N$.
Thus our calculations should provide us with an accurate picture
of what happens to bulk thermodynamics at $N=\infty$. For a more
detailed version of this work, which was conducted with
Dr.~M.~Teper, we refer to \cite{pressure_paper}.

\section{Lattice setup and methodology} \label{sec:lattice}

We define the gauge theory on a discretised periodic Euclidean four
dimensional space-time with $L^3_s\times L_t$ sites, and perform
Monte-Carlo simulations of a simple Wilson action. We use the
Kennedy-Pendelton heat bath algorithm for the link updates, followed
by five over-relaxations of all the $SU(2)$ subgroups of $SU(N)$. To
evaluate bulk thermodynamics there are several approaches one can
use. The ``differential method'' was used in the past, and is known
to produce the unphysical result of a negative pressure close to $T_c$.
The reason for that is the use of a perturbative beta function in a
regime where the coupling is not small. In our case, larger values of
$N$ drive us to smaller values of $L_t=5$, and therefore to a
lattice spacing of $a\simeq 1/(5T_c)$- too coarse for the differential
method. An additional possibility is the use of a direct evaluation
of the density of states. Unfortunately, modern methods like the Wang-Landau
algorithm did not converge for the case of $SU(8)$. This leaves us
with the ``integral method'' (see for example \cite{Boyd_et_al})
 in which the pressure $p$ and interaction measure $\Delta$ are
\begin{equation}
p/T^4
=
6 L^4_t \int^\beta_{\beta_0} d\beta^\prime
(\langle u_p \rangle_T - \langle u_p \rangle_0), \qquad
\label{eq:pint2}
\Delta/T^4
= 6 L_t^4
(\langle u_p(\beta) \rangle_0 - \langle u_p(\beta) \rangle_T)
\times
\frac{\partial \beta}{\partial \log (a(\beta))}.
\label{eq:Final_delta}
\end{equation}
Here $\langle u_p \rangle_{T}$ is the plaquette average on a $T> 0$
lattice with $L_t<L_s$, while $\langle u_p \rangle_{0}$ is measured
on a lattice with relatively large $L_t=L_s$. Let us mention that in
the expression for the pressure in \Eq{eq:Final_delta} we dropped
the pressure at temperature $T_0$
corresponding to $\beta_0$. To justify
this one usually takes $\beta_0$ much lower than the transition
coupling $\beta_c$, which means that the systematic error will be
very small.

 We performed calculations of $\langle u_p
\rangle_T$ in $SU(4)$ on $16^3 5$ lattices and in $SU(8)$ on $8^3 5$
lattices for a range of $\beta$ values corresponding to $T/T_c \in
[0.89,1.98]$ for $SU(4)$, and to $T/T_c\in [0.97,1.57]$ for $SU(8)$.
Since we use $L_t=5$, while the data for $SU(3)$ in
\cite{Boyd_et_al} is for $L_t=4,6,8$, we also performed simulations
for $SU(3)$ on $20^3 5$ lattices with $T/T_c\in [1,2]$. The measured
plaquette averages are presented in \cite{pressure_paper}.

We performed the `$T=0$' calculations of $\langle u_p\rangle_0$
on $20^4$ lattices for SU(3), and on
$16^4$ lattices for SU(4), which are known to be effectively at $T=0$
\cite{Lucini:2001ej,Lucini:2004my} for the couplings involved.
For SU(8)
however, using $8^4$ lattices would not be adequate for the
largest $\beta$-values, and we
take instead the SU(8) calculations on larger
lattices in
\cite{Lucini:2004my},
and interpolate between the values of $\beta$ used there with the
ansatz
$\langle u_p \rangle_0(\beta)
=\langle u_p \rangle^{P.T.}_0(\beta)
+\frac{\pi^2}{12}\frac{G_2}{N\sigma^2}(a\sqrt{\sigma})^4
+c_4g^8+c_5g^{10}, $
where $\langle u_p \rangle^{P.T.}_0(\beta)$ is the lattice
perturbative result to ${\cal O}(g^6)$ from
\cite{Alles:1998is}
and $N=8$.
Our best fit has $\chi^2/{\rm dof}=0.93$ with ${\rm dof}=2$,
and the best fit parameters are $c_4=-6.92$, $c_5=26.15$, and a
gluon condensate of $\frac{G_2}{N\sigma^2}= 0.72$.

\section{Finite volume effects} For $N=4,8$, one is able to use
lattice volumes much smaller than what one needs for $SU(3)$ (like
those in \cite{Boyd_et_al}) as the longest correlation length
decreases rapidly with $N$ \cite{Lucini:2003zr,Lucini:2005vg}. This
is also theoretically expected, much more generally, as $N\to\infty$
(see for example \cite{NN}). The main remaining concern has to do
with tunneling configurations, which occur only at $T_c$ when
$V\to\infty$. On our finite volumes, this is no longer true, and we
minimise finite-$V$ corrections by calculating the average
plaquettes only in field configurations that are confining, for
$T<T_c$, or deconfining, for $T>T_c$. For SU(3), where the phase
transition is only weakly first order, it is not practical to
attempt to separate phases. This will smear the apparent variation
of the pressure across $T_c$ in the case of SU(3).

To confirm that our finite volume effects are under control we have compared the SU(8) value of $\langle u_p(\beta) \rangle$ as
measured in the deconfined phase of the our $8^3\times 5$ lattice
with other $L_s^3\times 5$ results from other studies
\cite{Hagedorn_paper}, and find that the results are consistent
at the $2$ sigma level.
We perform a similar check for the confined phase on the same lattices, and again find that finite volume effects are small, mostly on a one sigma level.
A similar check for the confined phase on $L^4$ lattices leads to conclude that
a size $L=8$ in SU(8) is not
large enough, as we find that the plaquette average has a significant change (on a $16$ sigma level) from $L=8$ to $L=16$ for our largest value of $\beta$.
By contrast, for $SU(4)$ the finite volume effects seem not to be
large on the $16^4$ lattice as we checked for our largest value of
$\beta=11.30$. There the value of the plaquette on a $20^4$ lattice is
consistent within $\sim 2.3$ sigma with the value on a $16^4$
lattice.  The data supporting these checks, together with the checks related to the next paragraph, is presented in \cite{pressure_paper}.

\section{Finite lattice spacing corrections}
\label{finite_a}

To evaluate the scaling of the temperature $T$ with the coupling
$\beta=2N/g^2$, and the derivative in \Eq{eq:Final_delta}, we use
calculations of the string tension, $\sigma$, in lattice units (e.g.
\cite{Lucini:2005vg}). A more natural scale 
here is the transition temperature $T_c$. However this involves a
calculation of the transition coupling $\beta_c$ for many values of $L_t$, and
for several values of $L_s$, which is a very large scale project. Also the difference
between this scale fixing and the one according to $\sigma$ comes from the
$O(a^2)$ corrections of $T_c/\sqrt{\sigma}$. A check of the latter
for $SU(3),SU(8)$ shows that $T_c/\sqrt{\sigma}$ for $L_t=5,8$, are
the same within the errors. Indeed when we compare $T/T_c(\beta)$
from \cite{Boyd_et_al}, and as determined here for $N=3$, we
find that they fall on top of each other. This is not the case for
$SU(4)$, but there the $5\sigma$ difference in $T_c/\sqrt{\sigma}$ of
$L_t=5$, and $L_t=8$ is only at the level of 
$2\%$. As a result we may slightly overestimate $T/T_c$ when it is $\sim
8/5$ for this $N=4$. Finally the value of $T_c/\surd\sigma$ for the
different gauge groups and $L_t=5$ in taken from
\cite{Lucini:2003zr,Lucini:2005vg}.

In presenting our results for the pressure, we shall normalize to
the lattice Stephan-Boltzmann result given by $\left( p/T^4
\right)_{\rm{free-gas}}=(N^2-1)\frac{\pi^2}{45}\times R_I(L_t). $
Here $R_I$ includes the effects of discretization errors in the
integral method \cite{Engels:1999tk,pressure_paper}. The same
normalisation is applied for the interaction measure $\Delta$, and
for the energy density $\epsilon=\Delta+3p$, and entropy density
$s=(\Delta+4p)/T$.

\begin{figure}[htb]
\begin{center}
\includegraphics[width=6cm,clip]{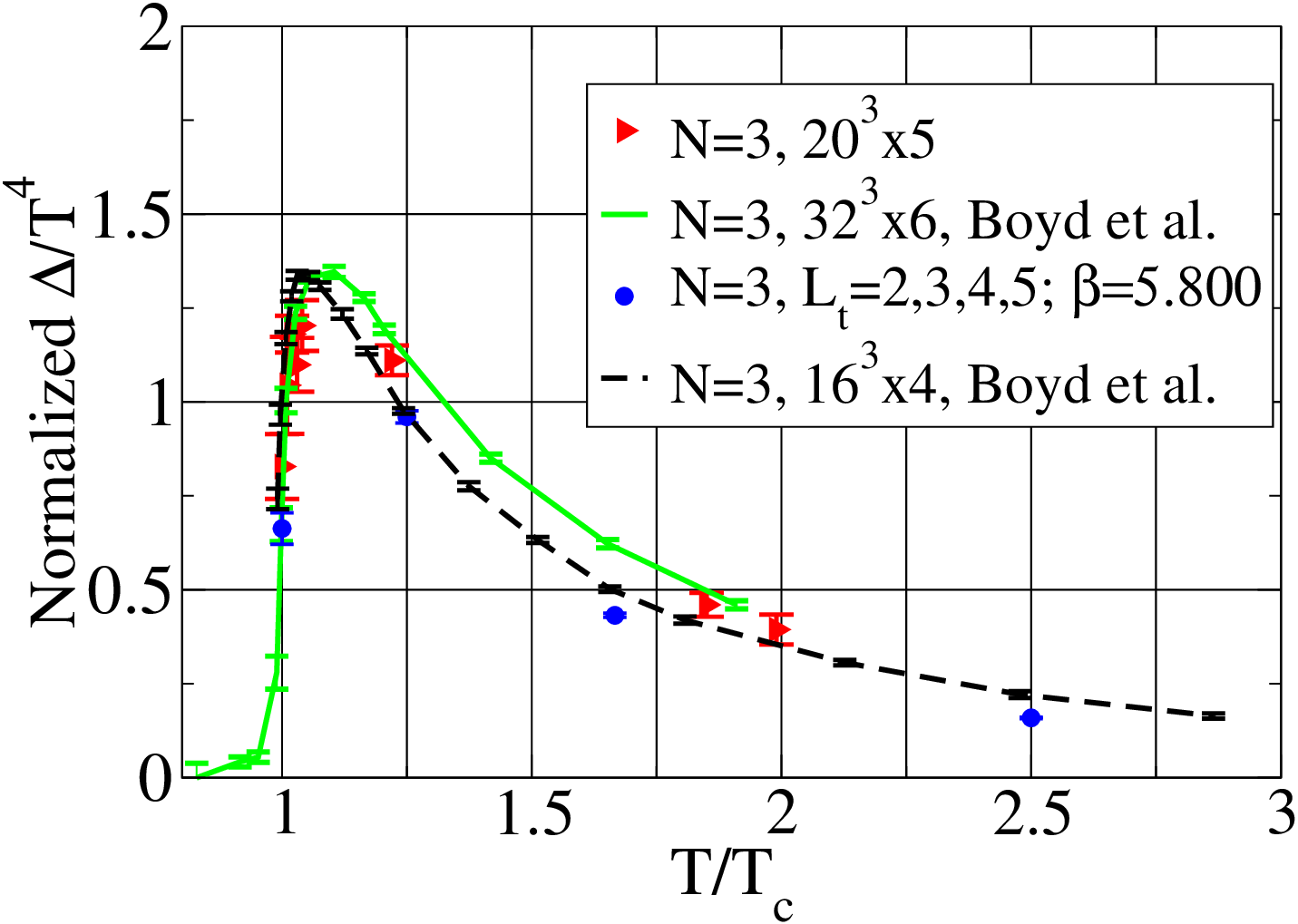} \qquad
\includegraphics[width=6cm,clip]{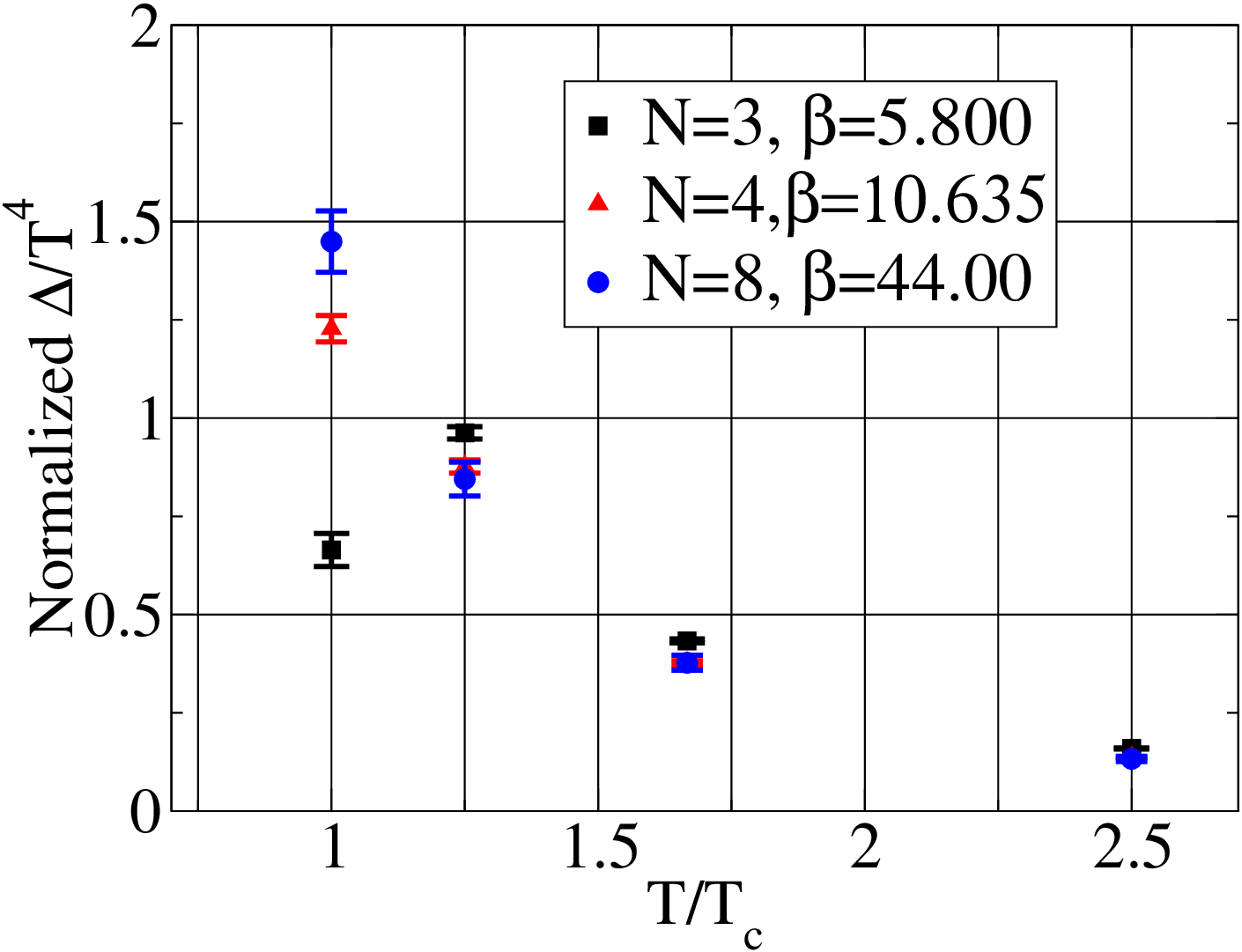}\caption{$\Delta$ for $N=3$ (on the
left), and $N=3,4,8$ (on the right) normalized to the lattice
Stephan-Boltzmann pressure, including the full discretization errors
given by $R_I$. The lines are the $SU(3)$ results and $L_t=4,6$ from
\cite{Boyd_et_al}.} \label{fig1} \vskip -0.5cm
\end{center}
\end{figure}

To check that we understand the systematics of finite lattice
spacing corrections when presenting the normalized results, we
compare $\Delta/T^4$ for $SU(3)$ with $L_t=4,6$ from
\cite{Boyd_et_al}, $L_t=5$ from the current study (in both these cases $T$ was increased by increasing
$\beta$), and $L_t=2,3,4$ from
\cite{Lucini:2005vg} (where $T$ was increased by decreasing $L_t$). The result of the
comparison is presented in the left panel of Fig.~\ref{fig1}, where
one indeed sees that the normalization by the Stephan-Boltzmann free
gas shows a systematic increase of $\Delta$ towards the continuum
limit.

\section{Results} \label{results}

Before we present our results for the bulk thermodynamical
quantities, let us note that as the ideas of reduction at large-$N$ (for example see the
recent systematic studies in \cite{NN}) have an interesting
realization here. 
By relabeling the axes, the authors of \cite{GN} have argued that at $N=\infty$ Wilson
loops should not change as a function of $T$ for $T<T_c$. This will
be reflected in our case in a $\<u_p\>_{T=T_c^-}$ which is very
similar to $\<u_p\>_0$. Indeed when we compare the two we find that
the difference is on a $\sim 2\sigma$ level for $SU(4),SU(8)$, while
 from the data of \cite{Engels:1999tk} it is on the $15\sigma$ level for
 $SU(3)$. This also means that the systematic error, ignored in
 \Eq{eq:Final_delta} is very small for the larger values of $N$.

\begin{figure}[htb]
\begin{center}
\includegraphics[width=7.1cm,clip]{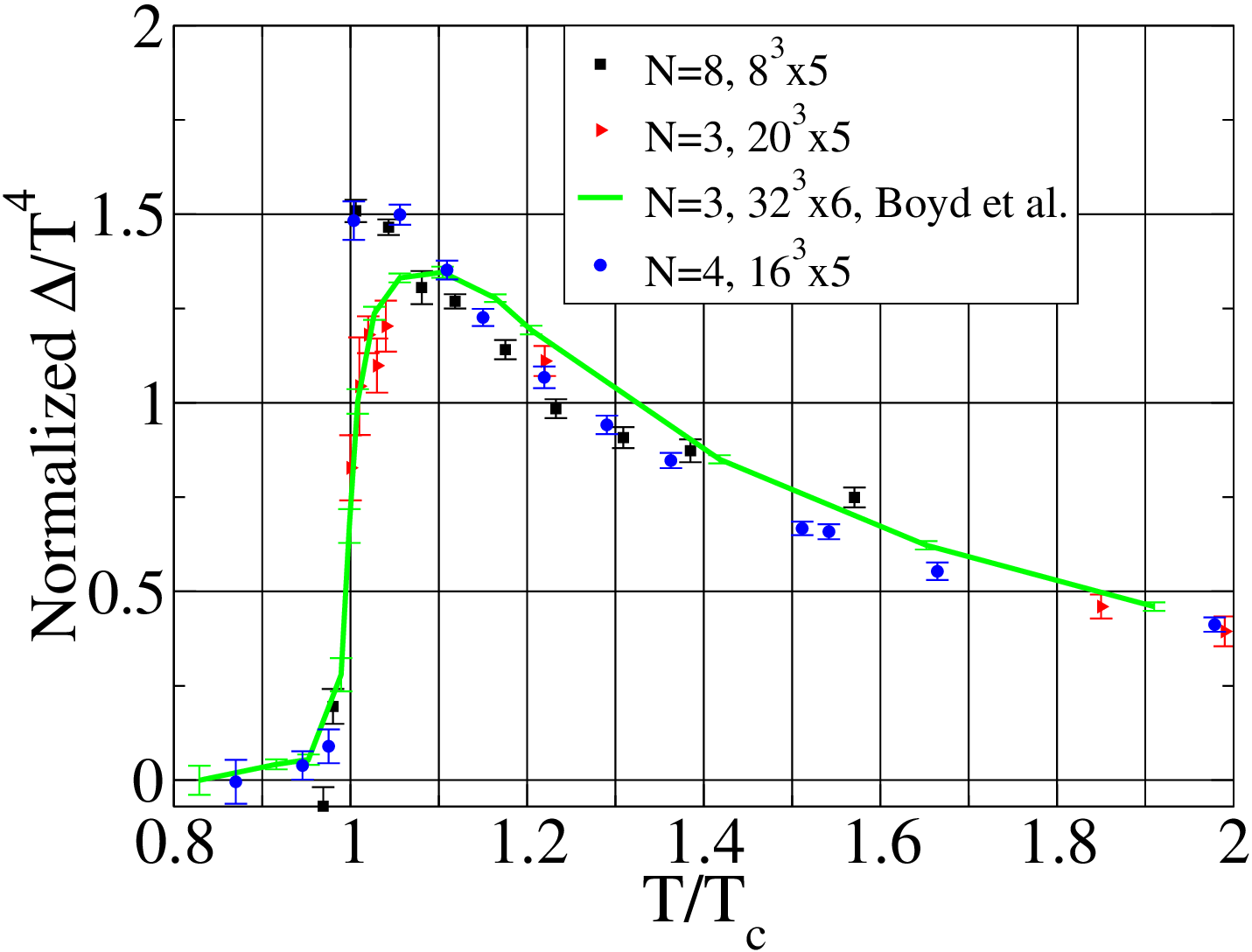}
\caption{$\Delta/T^4$ normalized to the lattice Stephan-Boltzmann
pressure, including the full discretization errors given by $R_I$,
for $N=3,4,8$. The line are the $SU(3)$ results and $L_t=6$ from
\cite{Boyd_et_al}}. \label{fig2} \vskip -0.5cm
\end{center}
\end{figure}

All the results we present are as a function of $T/T_c$. It is
therefore important to note that the dependence of $T_c$ on $N$ is
very weak, indeed like many other quantities in the pure gauge
theory \cite{Mikelattice05}.
 We present our $N=4$ and $N=8$ results for $\Delta/T^4$ in the right panel of
 Fig.~\ref{fig1}, and on Fig.~\ref{fig2}. In the former we change
 $L_t$ so to increase $T$ up to $T=2.5T_c$, while in the latter $T$ is
 increased by changing
 $\beta$. In both cases we find modest/small changes between the
 different groups. Nonetheless in the vicinity of $T_c$, the
 differences are large, presumably because the transition in $SU(3)$ is
 only weakly first order.

\begin{figure}[htb]
\begin{center}
\includegraphics[width=7.1cm,clip]{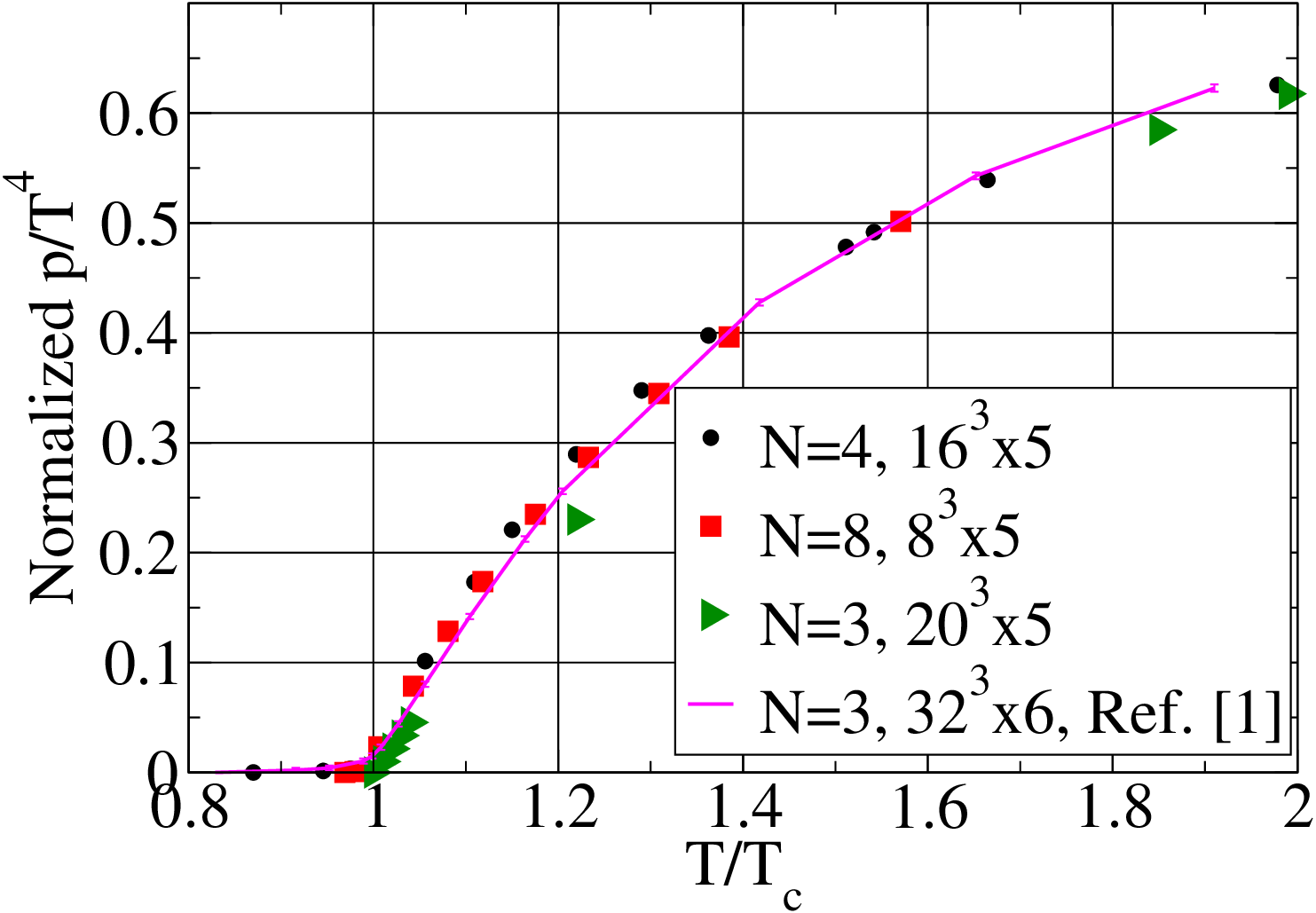} \qquad \includegraphics[width=7.1cm,clip]{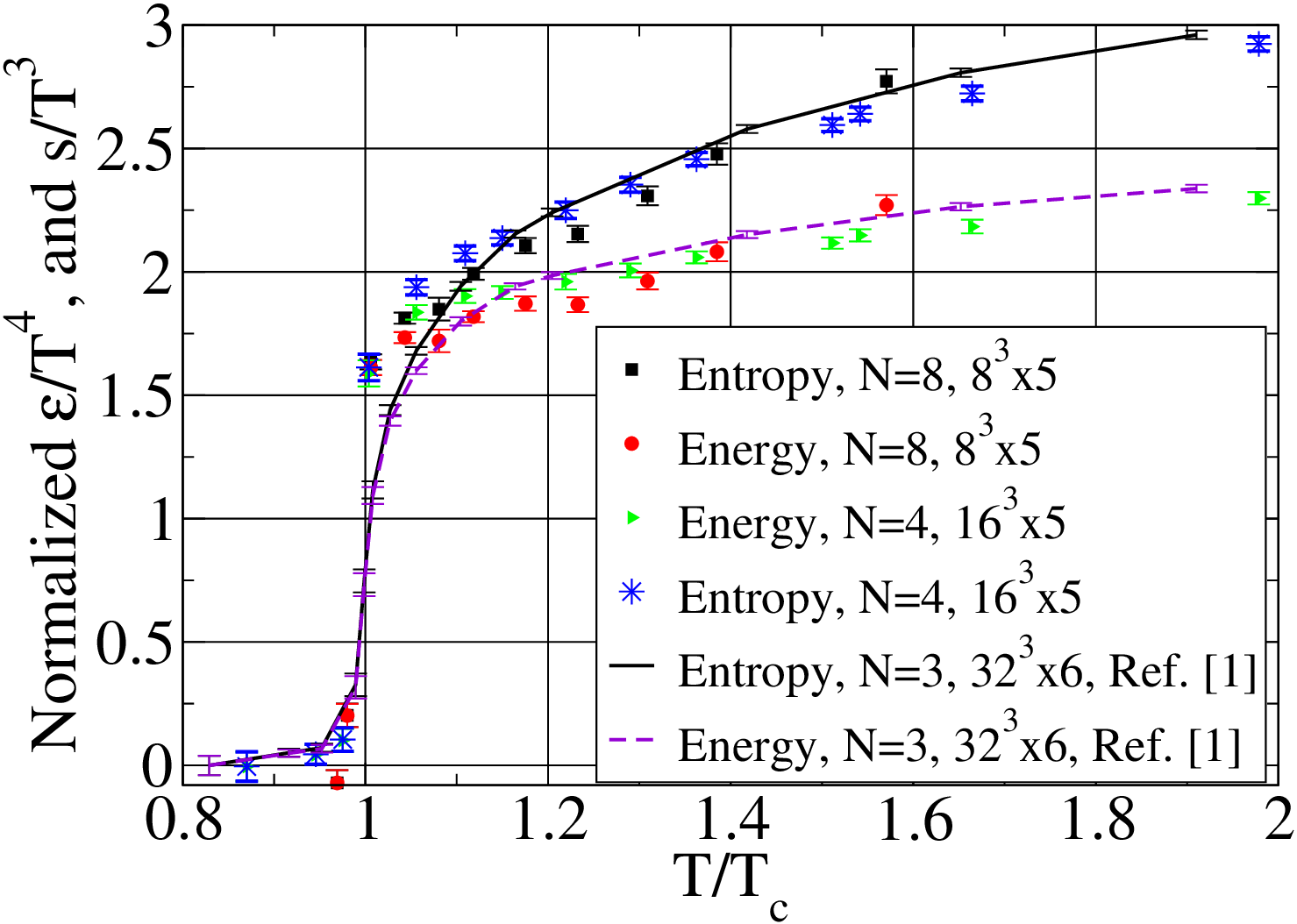}\caption{The pressure (on the left), and energy and entropy densities (on the right) normalized to the lattice Stephan-Boltzmann
pressure, including the full discretization errors given by $R_I$.
In the pressure plot, the symbols' vertical sizes are representing
the largest error bars (which are received for the highest
temperature). The lines are the $SU(3)$ results and $L_t=6$ from
\cite{Boyd_et_al}.} \label{fig3} \vskip -0.5cm
\end{center}
\end{figure}

  The plot of $p/T^4$ is presented in the left panel of Fig.~\ref{fig3}.
We also show our calculations of the $SU(3)$ pressure for $L_t=5$,
as well as the $L_t=6$ calculations from \cite{Boyd_et_al}. In the
right panel of Fig.~\ref{fig3} we present results for the normalized
energy density $\epsilon$, and normalized entropy density $s$. The
lines are the $SU(3)$ result of \cite{Boyd_et_al} with $L_t=6$.
Again we see very little dependence on the gauge group, implying
very similar curves for $N=\infty$.
One can clearly infer that the pressure in the $SU(4)$ and $SU(8)$
cases is remarkably close to that in $SU(3)$ and hence that the
well-known pressure deficit observed in SU(3) is in fact a property
of the large-$N$ planar theory. This implies that the dynamics that
drives the deconfined system far from its noninteracting gluon gas
limit, must remain equally important in the $N=\infty$ planar
theory. This is encouraging since this limit is simpler to approach
analytically, for example using gravity duals, and also because it
can serve to constraint and point to important ingredients of
analytical approaches. For example, in perturbation theory, it tells
us that the important contributions must be planar, and although the
current calculated contributions are indeed all planar, this is not guaranteed
at 
$O(g^7)$ \cite{York}. In models focusing on resonances and bound
states, it must be that the dominant states are coloured, since the
contribution of colour singlets will vanish as $N\to\infty$. Models
using `quasi-particles' should  place these in colour
representations that do not exclude their presence at $N=\infty$,
and in fact give them $T$-dependent properties which depend weakly
on $N$. Also, topological fluctuations should play no role in this
deficit since the evidence is that there are no topological
fluctuations of any size in the deconfined phase at large-$N$
\cite{Lucini:2004yh,DelDebbio:2004rw}.

\section*{Acknowledgments}
We are thankful to J.~Engels for discussions on the discretisation
errors, $R_I$, of the
 free lattice gas pressure, and for giving us the numerical routines to
 calculate them. We also thank the workshop
  organizers for the opportunity to present these work.


\begin{thebibliography}{99}

\bibitem{Heinz}
  U.~W.~Heinz,
  J.\ Phys.\ G {\bf 31}, S717 (2005)
  [arXiv:nucl-th/0412094].
  E.~V.~Shuryak,
  Nucl.\ Phys.\ A {\bf 750}, 64 (2005)
  [arXiv:hep-ph/0405066].


\bibitem{Boyd_et_al}
G.~Boyd, J.~Engels, F.~Karsch, E.~Laermann, C.~Legeland, M.~Lutgemeier
and B.~Petersson, 
 Nucl.\ Phys.\ B {\bf 469}, 419 (1996) [arXiv:hep-lat/9602007].

\bibitem{Petreczky:2004xs}
  P.~Petreczky,
  Nucl.\ Phys.\ Proc.\ Suppl.\  {\bf 140}, 78 (2005)
  [arXiv:hep-lat/0409139].

\bibitem{Philippe}
  P.~de Forcrand,
  these proceedings.

\bibitem{NN}
  R.~Narayanan and H.~Neuberger,
  arXiv:hep-lat/0509014.

\bibitem{Lucini:2002ku}
B.~Lucini, M.~Teper and U.~Wenger, 
 Phys.\ Lett.\ B {\bf 545}, 197 (2002) [arXiv:hep-lat/0206029].

\bibitem{Lucini:2003zr}
B.~Lucini, M.~Teper and U.~Wenger, 
 JHEP {\bf 0401}, 061 (2004) [arXiv:hep-lat/0307017].

\bibitem{Lucini:2005vg}
B.~Lucini, M.~Teper and U.~Wenger, 
JHEP {\bf 0502}, 033 (2005) [arXiv:hep-lat/0502003].

\bibitem{Lucini:2001ej}
B.~Lucini and M.~Teper, 
 JHEP {\bf 0106}, 050 (2001) [arXiv:hep-lat/0103027].

\bibitem{Lucini:2004my}
B.~Lucini, M.~Teper and U.~Wenger, 
JHEP {\bf 0406}, 012 (2004) [arXiv:hep-lat/0404008].

\bibitem{pressure_paper}
  B.~Bringoltz and M.~Teper,
  Phys.\ Lett.\ B {\bf 628}, 113 (2005)
  [arXiv:hep-lat/0506034].


\bibitem{Alles:1998is}
B.~Alles, A.~Feo and H.~Panagopoulos, 
Phys.\ Lett.\ B {\bf 426}, 361 (1998) [Erratum-ibid.\ B {\bf 553}, 337 (2003)] [arXiv:hep-lat/9801003].

\bibitem{Hagedorn_paper}
B.~Bringoltz and M.~Teper,
 arXiv:hep-lat/0508021.

\bibitem{Engels:1999tk}
J.~Engels, F.~Karsch and T.~Scheideler,
Nucl.\ Phys.\ B {\bf 564}, 303 (2000) [arXiv:hep-lat/9905002].
J.~Engels,
private Communications (2005)

\bibitem{GN}
  A.~Gocksch and F.~Neri,
  Phys.\ Rev.\ Lett.\  {\bf 50}, 1099 (1983).

\bibitem{Mikelattice05}
  M.~Teper,
  arXiv:hep-lat/0509019.

\bibitem{York}
  Y.~Schroder,
  private communications (2005).

\bibitem{Lucini:2004yh}
  B.~Lucini, M.~Teper and U.~Wenger,
  Nucl.\ Phys.\ B {\bf 715}, 461 (2005)
  [arXiv:hep-lat/0401028].

\bibitem{DelDebbio:2004rw}
  L.~Del Debbio, H.~Panagopoulos and E.~Vicari,
  JHEP {\bf 0409}, 028 (2004)
  [arXiv:hep-th/0407068].


\end{thebibliography}
\end{document}